\documentclass[aps,prl,floatfix,superscriptaddress,twocolumn,footinbib]{revtex4-1}
\usepackage{graphicx}
\usepackage{dcolumn}
\usepackage{bm}
\usepackage{makecell}
\usepackage{color}

\begin{document}

\title{Beyond no-go theorem' Weyl phonons}

\author{Qing-Bo Liu}
\author{Xiang-Feng Yang}%
\author{Ao Lou}%
 \affiliation{School of Physics and Wuhan National High Magnetic Field Center,
Huazhong University of Science and Technology, Wuhan 430074, People's Republic of China.}

\author{Hua-Hua Fu}
\altaffiliation{hhfu@hust.edu.cn}
\affiliation{School of Physics and Wuhan National High Magnetic Field Center,
Huazhong University of Science and Technology, Wuhan 430074, People's Republic of China.}
\affiliation{Institute for Quantum Science and Engineering,
 Huazhong University of Science and Technology, Wuhan, Hubei 430074, China.}

\date{\today}

\begin{abstract}
By using \emph{ab initio} calculations and symmetry analysis, we define a new class of Weyl phonons, i.e., isolated Weyl phonons (IWPs), which are characterized by Chern number $\pm$2 or $\pm$4 in their acoustic phononic spectra and protected by the time inversion symmetry and point group symmetries. More importantly, their particular topological feature make them circumvent from the no-go theorem. Some high-symmetry points, behaving as isolated Weyl points in the space groups (SGs) of the related phononic systems, tend to form IWPs. As enumerated in Table I, the IWPs are located at the center of three-dimensional Brillouin zone (BZ), and protected by the time-reversal symmetry ($\cal T$) and the corresponding point group symmetries. Moreover, a realistic chiral crystal material example of K$_2$Mg$_2$O$_3$ in SG 96, a monopole IWP with Chern number -2 is found at the high-symmetry point $\Gamma$, and in another material example of Nb$_3$Al$_2$N in SG 213, a monopole IWP with Chern number +4 is confirmed at the point $\Gamma$. It is interesting that that IWPs can not form the surface arcs in the surface BZ, which has not been reported in the phononic systems to present. Our theoretical results not only uncover a new class of Weyl phonons (IWPs), but also put forwards an effective way to search the IWPs in spinless systems.

\end{abstract}

\maketitle

{\color{blue}\emph{Introduction.}} Weyl points (WPs) are named after Hermann Weyl {\color{blue}\cite{1}}, who firstly proposed them as more elementary particles than Dirac points to preserve the Lorentz symmetry, since a pair of WPs with opposite chirality can be hybridized to compose a Dirac point. It is well established that WPs are divided into two categories, i.e., conventional WPs and unconventional WPs {\color{blue}\cite{2,3,4,5,6,7}}. The former are contributed by two degenerated bands with a monopole charge of $\pm$1, and need the protection of translation symmetry in materials. While the latter are contributed by a quantized monopole charge equal to or larger than one {\color{blue}\cite{8}}, and require the protections of additional crystalline symmetries, such as $C_{31}$ {\color{blue}\cite{9}}, $S_4$ symmetries {\color{blue}\cite{9}} or others. According to the no-go theorem, the compensation effects should be satisfied in Weyl systems. In other words, the total charge of WPs should be zero. For example, Weng {\color{blue}\cite{10}} and Xu et al. {\color{blue}\cite{11}} have observed many pair of WPs in TaAs in the experiment in the electronic systems; spin-1 Weyl phonon with Chern numbers of 0 and $\pm$2 and charge-2 Dirac phonons with Chern number of $\pm$2 have been predicted in theory and observed in the experiment in phononic systems {\color{blue}\cite{12,13}}, which can form the double-helicoid surface states and large surface arcs. At the same time, a charge-four Weyl phonon with the charge of +4 can be compensated with four single-Weyl phonons with the Chern number -1 in BiIrSe in SG 198 {\color{blue}\cite{14}}, which can form quadratic-helicoid surface states and quadratic-helicoid surface acrs. Thus, many interesting physical phenomena in the Weyl systems drive us to explore a new class of Weyl phonons in phononic systems.

On the other side, searching for the Dirac, Weyl and nodal lines phonons has been another topic in the filed of topological phonons {\color{blue}\cite{15,16,17,18,19,19_1,19_2,19_3,19_4,19_5}}, because phonons, as a typical kind of bosons, don't satisfy the Pauli exclusion principle. Moreover, topological phonons can exhibit many special features, such as phonon diode effect {\color{blue}\cite{20}}, the phonon Hall effect {\color{blue}\cite{21}} and so on. To our best knowledge, Dirac phonons have been classified in 230 SGs {\color{blue}\cite{18}}, and been reported in the high-pressure CuCl {\color{blue}\cite{17}} in the past year. At the same time, the nodal lines have been classified into various categories, including isolated closed nodal rings {\color{blue}\cite{22,23,24,28}}, nodal chain {\color{blue}\cite{26}}, nodal nets {\color{blue}\cite{26_1}}, nodal links {\color{blue}\cite{17,26}}, nodal knots {\color{blue}\cite{27_1,27_2}}, straight nodal lines {\color{blue}\cite{19_5}}, and opened nodal lines {\color{blue}\cite{17}}. As for Weyl phonons, single Weyl phonons with the charge of $\pm$1 {\color{blue}\cite{29}}, charge-two Weyl phonons with Chern number of $\pm$2 {\color{blue}\cite{15}} and charge-four Weyl phonons with the charge of $\pm$4 {\color{blue}\cite{14}} have all been reported successively in the phononic systems, which all induce inspiring surface arcs in the surface BZ and obey the no-go theorem. One may wonder that does it exist isolated Weyl phonons (IWPs) which are beyond no-go theorem and thus do not form the surface arcs? If exist, two questions will arise naturally: (i) how do IWPs exist in all 230 space groups (SGs)? and (ii), what are the topological nontrivial features of IWPs in realistic materials?

\begin{figure}
\includegraphics[width=3.20in]{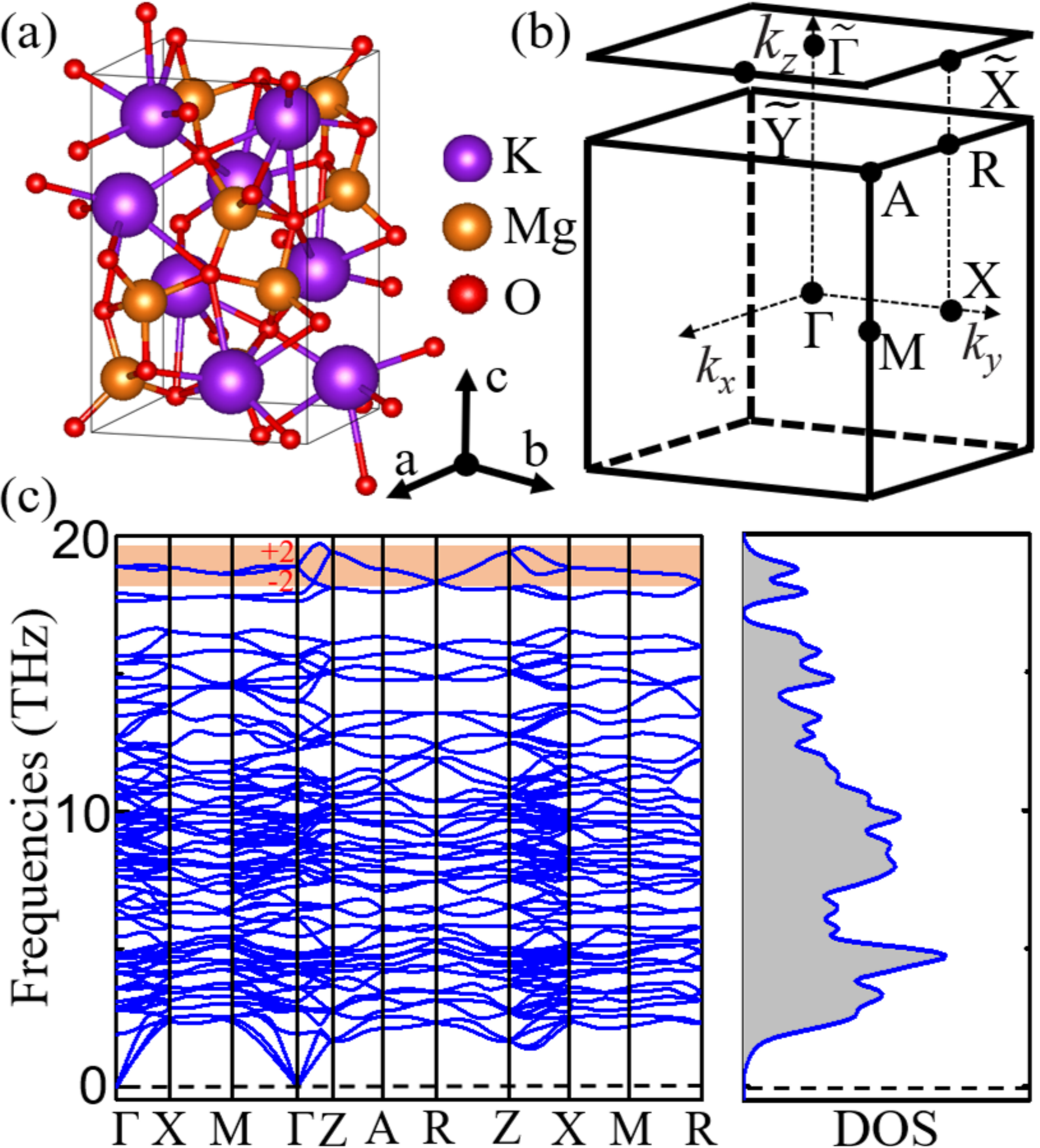}
\caption{The crystalline structures and phonon dispersions of K$_2$Mg$_2$O$_3$ in SG 96. (a) Crystal structure of K$_2$Mg$_2$O$_3$ in a primitive cell, where the purple (orange and red) atoms stand for K (Mg and O). (b) The bulk BZ of K$_2$Mg$_2$O$_3$ and the (001) surface BZ. (c) The phonon spectra of K$_2$Mg$_2$O$_3$ along high-symmetry lines and a red boxes is centered around IWP at $\Gamma$ point, and the Chern numbers ($\pm$2) of some nontrivial phonon branches around the $\Gamma$ point. (d) The phononic density of states (DOS) of K$_2$Mg$_2$O$_3$.}
\end{figure}

In this work, by checking the symmetries of 230 SGs, we exhaust the all IWPs existing in the 3D Brillouin zone (BZ) of bosonic systems, and enumerate them at the high-symmetry points of SGs in Table~\uppercase\expandafter{\romannumeral1}, in which we find that the phononic bands tend to be doubly degenerate at those high-symmetry points and every twofold degeneracy maybe represents a IWP. To confirm our findings, we study further the phononic spectra and the corresponding surface states of two realistic crystal material examples, \emph{i.e}., K$_2$Mg$_2$O$_3$ with Chern number of $\pm$2 at the $\Gamma$ point in SG 96 and Nb$_3$Al$_2$N with charge of $\pm$4 at the $\Gamma$ point in SG 213, by performing the first-principles calculations, we find that in the former, a charge-two IWP (CTIWP) appear at the point $\Gamma$ with the Chern number of -2, which are formed by the 83$^{rd}$ and 84$^{th}$ phonon bands in the first BZ, while in the latter, a charge-four IWP (CFIWP) appears at the point $\Gamma$ with the Chern number of +4, which are formed by the 59$^{th}$ and 60$^{th}$ phononic bands in the first 3D BZ. The CTIWPs and CFIWPs in both materials are monopole. Moreover, the CTIWPs (CFIWPs) can not induce double (quadruple)-helicoid surface states and noncontractable surface arcs, only form quadratic Dirac cones in the surface and isolated points in the isofrequency surface contours. Our theoretical results not only put forwards an effective route to search the all IWPs in bosonic systems, but also provide a platform to study topological phonons beyond no-go theorem' Weyl phonons.

\begin{table*}[!t]
\caption{The completed list of circumventing no-go theorem' Weyl phonons. The first column indicates the SG number, the second column indicates the SG symbol, the third column shows the corresponding high-symmetry points. The fourth column shows the point groups (PGs), the fifth column shows the Chern numbers (CNs), the sixth column shows the irreducible representations (Irreps), the seventh column shows the Generators and the eighth column shows the realistic thermodynamically stable material examples.
}\label{tab:CFWPs}
\begin{tabular}{p{1.5cm}p{2cm}p{1.5cm}p{1cm}p{1cm}p{1cm}p{3cm}p{5.8cm}}
\hline
\hline
\makecell[c]{SG No.} &\makecell[c]{SG Symbol} &\makecell[c]{$k$-point} &\makecell[c]{PGs} &\makecell[c]{CNs}  &\makecell[c]{Irreps} &\makecell[c]{Generators}  &\makecell[c]{Thermodynamically Stable Materials} \\
\hline
\makecell[c]{92} & \makecell[c]{$P4_{1}2_{1}2$}  &\makecell[c]{$\Gamma$} &\makecell[c]{$D_{4}$}   &\makecell[c]{$\pm$2} & \makecell[c]{$\Gamma_5$} & \makecell[c]{$\{C_{4z}^+|00\frac{1}{4}\}$ \\ $\{C_{2x}|\frac{1}{2}\frac{1}{2}0\}$, $\mathcal{T}$} & \makecell[c]{$BaPt_{2}S_{3}, CdP_{2}, GeO_{2}, LiAlO_{2}$ \\ $ Pt_{3}I_{8}, Rb_{2}Be_{2}O_{3}, SiO_{2}, ZnP_{2}$} \\
\hline
\makecell[c]{96} & \makecell[c]{$P4_{3}2_{1}2$} &\makecell[c]{$\Gamma$} &\makecell[c]{$D_{4}$} &\makecell[c]{$\pm$2} & \makecell[c]{$\Gamma_5$} & \makecell[c]{$\{C_{4z}^+|00\frac{3}{4}\}$ \\ $\{C_{2x}|\frac{1}{2}\frac{1}{2}0\}$,$\mathcal{T}$} & \makecell[c]{$Ag_{2}HgO_{2}, CdP_{2}, K_{2}Mg_{2}O_{3}$ \\ $MgAs_{4}, Na_{2}Zn_{2}O_{3}, SiO_{2}, ZnP_{2}$ }\\
\hline
\makecell[c]{198} & \makecell[c]{$P2_13$} &\makecell[c]{$\Gamma$} &\makecell[c]{$T$} &\makecell[c]{$\pm$4} & \makecell[c]{$\Gamma_2$$\Gamma_3$} & \makecell[c]{$ \{C_{31}^+|000\}$\\$\{C_{2z}|\frac{1}{2}0\frac{1}{2}\}$\\$\{C_{2y}|0\frac{1}{2}\frac{1}{2}\}$,$\mathcal{T}$} & \makecell[c]{$BaAsPt, BeAu, BiRhSe, K_{3}SbO_{3}$\\  $SbIrS, SiRu, ZrSb, BaPPt$\\ $ BiTeIr, BiTePt, CoAsS, HfSO $\\ $ KMgBO_{3}$, $SiOs$, $SrMgNiH_{4}$, $ZrSeO$} \\
\hline
\makecell[c]{212} & \makecell[c]{$P4_332$} &\makecell[c]{$\Gamma$}  &\makecell[c]{$O$} &\makecell[c]{$\pm$4} & \makecell[c]{$\Gamma_3$} & \makecell[c]{$\{C_{31}^-|000\}$ \\ $\{C_{2z}|\frac{1}{2}0\frac{1}{2}\}$ \\ $\{C_{2x}|\frac{1}{2}\frac{1}{2}0\}$ \\ $\{C_{2a}|\frac{1}{4}\frac{3}{4}\frac{3}{4}\}$,$\mathcal{T}$} & \makecell[c]{$SrSi_{2}, BaGe_{2}$, $LiAl_{5}O_{8}$}\\
\hline
\makecell[c]{213} & \makecell[c]{$P4_132$}  &\makecell[c]{$\Gamma$} &\makecell[c]{$O$} &\makecell[c]{$\pm$4} & \makecell[c]{$\Gamma_3$} & \makecell[c]{$\{C_{31}^-|000\}$ \\ $\{C_{2z}|\frac{1}{2}0\frac{1}{2}\}$ \\ $\{C_{2x}|\frac{1}{2}\frac{1}{2}0\}$ \\ $\{C_{2a}|\frac{3}{4}\frac{1}{4}\frac{1}{4}\}$,$\mathcal{T}$} & \makecell[c]{$Na_{4}Sn_{3}O_{8}$, $CsBe_{2}F_{5}$, $Mg_{3}Ru_{2}$ \\ $Nb_{3}Al_{2}N, Nb_{3}Al_{2}C, V_{3}Zn_{2}N$ \\ $Mo_{3}Pd_{2}N$, $Mo_{3}Ni_{2}N, W_{3}Pd_{2}N$ \\ $W_{3}Ni_{2}N$, $V_{3}Ga_{2}N$ \\ $Ta_{3}Al_{2}C$, $Mo_{3}Co_{2}N$}\\
\hline
\hline
\end{tabular}
\end{table*}

{\color{blue}\emph{Symmetry analysis on IWPs with Chern numbers $\pm$2 and $\pm$4.}} To search the twofold IWPs, we should firstly find out the 2D irreducible representations (irreps) of the little groups (LGs) in the presence of time-reversal symmetry ($\mathcal{T}$). This searching covers the all 230 SGs and the lattice symmetries of material candidates should be located at the high-symmetry \emph{k}-points in the 3D BZ. Here, a 2D irrep stands for a twofold point at the high-symmetry point. Secondly, we exclude the SGs that contain the space-inversion symmetry ($\mathcal{P}$) to ensure that there no any nodal line is located at these high-symmetry points. Thirdly, we need to find the SGs which possess three two-fold skew axial symmetries along the $x$, $y$ and $z$ directions, which can form three nodal surfaces at the planes of $k_x=\pm\pi$, $k_y=\pm\pi$ and $k_z=\pm\pi$ in the first BZ. Finally, we screen out the twofold points at the high-symmetry points, and then calculate the charge number of these WPs one by one to gain the IWPs hosting the charge of $\pm$2 or $\pm 4$. Following this process, the all WPs meeting the requirements of IWPs are obtained as listed in Table~\uppercase\expandafter{\romannumeral1}.

Let us consider a two-band model of IWPs with Chern numbers $\pm$2 or $\pm$4 at the $\Gamma$ point in SGs 96 and 213. We find that the $\Gamma$ point in both 96 and 213 SGs possesses $\mathcal{T}$, thus we can derive the two-band $k\cdot{p}$ Hamiltonian of IWPs under $\mathcal{T}$ firstly. Based on this symmetry, the two-band $k\cdot{p}$ Hamiltonian of IWPs can be given as
\begin{equation}
\mathcal{H}_{kp}=g_x(k)\sigma_x+g_y(k)\sigma_y+g_z(k)\sigma_z,
\end{equation}
where $k=(k_x,k_y,k_z)$, $\sigma_{x,y,z}$ represents the three Pauli matrices, and $g_{x,y,z}(k)$ represents the complex functions $k_x$, $k_y$ and $k_z$. In the momentum space, $\mathcal{T}$ gives the following relation
\begin{equation}
\emph{T}\mathcal{H}_{kp}(k)\emph{T}^{-1}=\mathcal{H}_{kp}(-k),
\end{equation}
where \emph{T} is a complex conjugate operator under the 2D irreps, i.e., $\Gamma_5$ in SG 96 and $\Gamma_3$ in SG 213. Under the \emph{T} operation, $g_x(k)$ and $g_z(k)$ are even functions, while $g_y(k)$ is an odd one. Thus, the expanded $k\cdot{p}$ Hamiltonian as a function of $k$ in low-energy region can be described as
\begin{eqnarray*}
&g_x(k)&=\sum\limits_{i,j=x,y,z}a_x^{ij}k_{i}k_{j},   \\
&g_y(k)&=\sum\limits_{i=x,y,z}a_{y1}^ik_{i}+\sum\limits_{i,j,m=x,y,z}a_{y2}^{ijm}k_{i}k_{j}k_{m},   \\
&g_z(k)&=\sum\limits_{i,j=x,y,z}a_z^{ij}k_{i}k_{j}.
\end{eqnarray*}
Then, we consider further other required symmetry operations in the SGs 96 and 213. Here, the representation matrixes for $\{C_{4z}^+|00\frac{3}{4}\}$ ($C_{4z}^+$:$xyz \mapsto \bar y xz$) and $\{C_{2x}|\frac{1}{2}\frac{1}{2}0\}$ at $\Gamma$ point in SG 96  can be written as
\begin{eqnarray*}
&C_{4z}^+&=\left[\begin{array}{ll}
0 & -1 \\
1 & 0
\end{array}\right] =-i\sigma_y, \\
&C_{2x}&=\left[\begin{array}{ll}
 0 & -1 \\
-1 &  0
\end{array}\right]= -\sigma_x.\\
\end{eqnarray*}
Under the above two operations, the final $k\cdot p$-invariant Hamiltonian with the charges of $\pm$2 can be derived as,
\begin{eqnarray*}
\mathcal{H}_{kp}=a_{x}^{xx}\left(k_{x}^{2}-k_{y}^{2}\right)\sigma_{x}+a_{y1}^{z}k_{z}\sigma_{y},
\end{eqnarray*}
where $a_{x}^{xx}$ and $a_{y1}^{z}$ are constant coefficients. According to the above two-band $k\cdot{p}$ Hamiltonian, we conclude that the dispersions display a $k$-type feature along the [001] direction, while a $k^2$-type feature along other directions in the phononic bands. It should be stressed that this $k\cdot{p}$ model is also applicable at the $\Gamma$ point of the SG 92.

Finally, the representation matrixes for $\{C_{31}^-|000\}$ ($C_{31}^-$:$xyz \mapsto yzx$), $\{C_{2x}|\frac{1}{2}0\frac{1}{2}\}$, $\{C_{2z}|\frac{1}{2}0\frac{1}{2}\}$ and $\{C_{2a}|\frac{3}{4}\frac{1}{4}\frac{1}{4}\}$ ($C_{2a}$:$xyz \mapsto yx\bar z$) at $\Gamma$ point in SG 213 can be written as
\begin{eqnarray*}
&C_{31}^-&=\left[\begin{array}{ll}
-\frac{1}{2} & \frac{\sqrt{3}}{2} \\
-\frac{\sqrt{3}}{2} & -\frac{1}{2}
\end{array}\right], \\
&C_{2x}&=\left[\begin{array}{ll}
1 & 0 \\
0 & 1
\end{array}\right], C_{2x}=C_{2z},\\
&C_{2a}&=\left[\begin{array}{ll}
0 & 1 \\
1 & 0
\end{array}\right] = \sigma_x,\\
\end{eqnarray*}
Under the above three operations, the final $k\cdot p$-invariant Hamiltonian can be written as,
\begin{eqnarray*}
\mathcal{H}_{kp}=a_{x}^{xx}\left(k_{x}^{2}-k_{y}^{2}\right)\sigma_{x}+a_{y2}^{xyz}k_{x}k_{y}k_{z}\sigma_{y} \\
+\frac{1}{\sqrt{3}}a_{x}^{xx}\left(k_{x}^{2}+k_{y}^{2}-2k_{z}^{2}\right)\sigma_{z},
\end{eqnarray*}
where $a_{x}^{xx}$ and $a_{y2}^{xyz}$ are constant coefficients. According to the above two-band $k\cdot{p}$ Hamiltonian, we conclude that the dispersions display a $k^3$-type feature along the [111] direction, while a $k^2$-type feature along other directions in the phonon bands. It should be stressed that this $k\cdot{p}$ model is also applicable at $\Gamma$ points of 198 and 212 SGs.

\begin{figure*}
\includegraphics[width=17.5 cm]{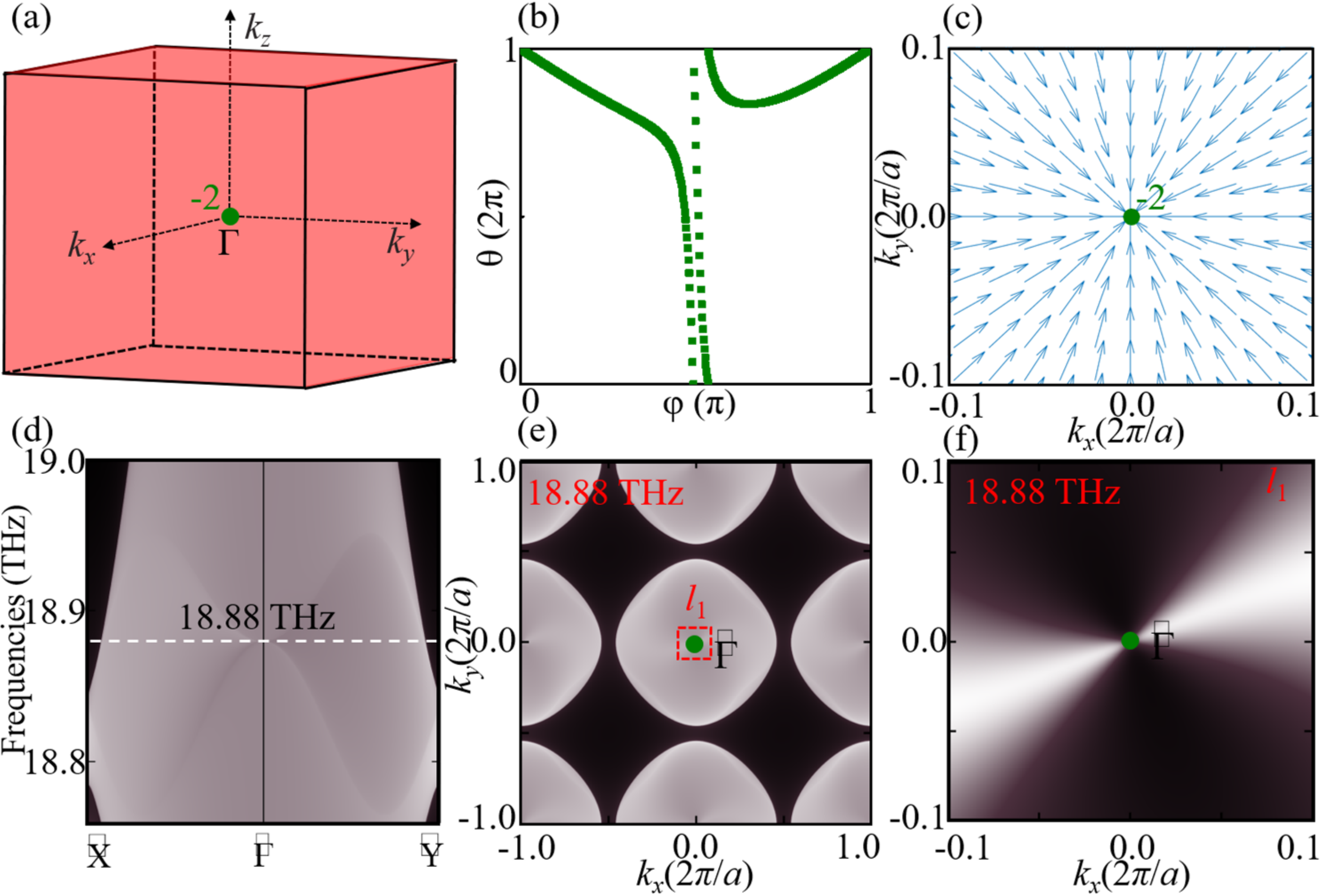}
\caption{The topological features of K$_2$Mg$_2$O$_3$ in SG 96. (a) The nodal wall and IWP in the first 3D BZ. (b) The evolutions of Wannier centers of the 83rd phonon band for the point $\Gamma$. (c) The distribution of the Berry curvature in the (001) plane. (d) The surface states of K$_2$Mg$_2$O$_3$ in the (001) surface BZ along the $\widetilde{\textrm{X}}$-$\widetilde{\textrm{$\Gamma$}}$-$\widetilde{\textrm{Y}}$. (e) and (f) The isofrequency surface contours at 18.88 THz.}
\end{figure*}

{\color{blue}\emph{Crystal structure and calculation methods.}} To confirm the existence of IWPs in realistic materials and to examine further their topologically nontrivial features, we turn to study the phononic spectra of two material samples, \emph{i.e.}, K$_2$Mg$_2$O$_3$ in SG $P4_{3}2_{1}2$ (No. 96) and Nb$_3$Al$_2$N in SG $P4_132$ (No. 213). We obtain the crystallographic data of K$_2$Mg$_2$O$_3$ and Nb$_3$Al$_2$N from Ref.{\color{blue}\cite{30}}, and their primitive cells are illustrated in Fig. 1(a) and 3(a), respectively. The primitive cell of K$_2$Mg$_2$O$_3$ contains twenty-eight atoms with eight K, and eight Mg and twelve O atoms, denoted by the purple, orange and red balls, where $a=b=6.490537$ $\textrm{{\AA}}$, $c=10.393529$ $\textrm{{\AA}}$ and the Wyckoff positions $8b$ (0.93287, 0.73674, -0.13792) for K atoms, $8b$ (0.92717, 0.77770, 0.38271) for Mg atoms, $8b$ (0.0.97139, 0.26443, 0.06470) for O atoms and $4a$ (0.38317, 0.38317, 0.5) for O atoms. The corresponding BZ is drawn in Fig. 1(b), where the black square is the (001) surface BZ. The primitive cell of Nb$_3$Al$_2$N contains twenty-four atoms with twelve Nb, eight Al and four N atoms, denoted by the green, blue and solid balls in Fig. 3(a), Where $a=b=c=7.079402$ $\textrm{{\AA}}$ and the Wyckoff positions $12d$ (0.7955, 0.0.9545, 0.375) for Nb atoms, $8c$ (0.17918, 0.82082, 0.0.52082) for Al atoms, and $4a$ (0.875, 0.125, 0.625) for N atoms. And the corresponding BZ is drawn in Fig. 3(b), where the blue square denotes the (001) surface BZ.

The phononic dispersions of all material examples are calculated by the density functional theory (DFT) using the Vienna \emph{ab initio} Simulation Package (VASP) with the generalized gradient approximation (GGA) in the form of Perdew-Burke-Ernzerhof (PBE) function for the exchange-correlation potential {\color{blue}\cite{31,32,33}}. An accurate optimization of structural parameters is employed by minimizing the interionic forces less than 0.001 $\textrm{eV}/\textrm{{\AA}}$ and an energy cut off at 500 eV. The BZ is gridded with 3$\times$3$\times$3 \emph{k} points. Then the phononic spectra are gained using the density functional perturbation theory (DFPT), implemented in the Phonopy Package {\color{blue}\cite{34}}. The force constants are calculated using a $2\times2\times2$ supercell. To reveal the topological nature of phonons, we construct the phononic Hamiltonian of tight-binding (TB) model and the surface local density of states (LDOS) with the open-source software Wanniertools code {\color{blue}\cite{35}} and surface Green's functions {\color{blue}\cite{36}}. The Chern numbers or topological charge of charge-two and charge-four WPs are calculated by Wilson loop method {\color{blue}\cite{37}} and the crystal structures are obtained from Materials Project (MP) {\color{blue}\cite{30}}. The irreps of the phonon bands are computed by the program {\color{blue}$ir2tb$} on the phononic Hamiltonin of TB model {\color{blue}\cite{38}}.

{\color{blue}\emph{Topological features and exotic surface states of charge-two IWPs.}} We divide the IWPs into two types, i.e., charge-two IWPs and charge-four IWPs in the Table-I, and we will discuss the topological features of charge-two isolated Weyl phonons (CTIWPs) firstly. The phononic bands and the corresponding phonon density of states (PDOS) of K$_2$Mg$_2$O$_3$ are drawn in Fig. 1(c) and 1(d), where no imaginary frequencies indicate that this material is thermodynamically stable. From which, one can see that there are two band crossings located at the point $\Gamma$ and along the high-symmetry lines X-M, Z-A-R and M-R in the optical branches. Here, we mainly discuss a perfect CTIWP which is formed by the 83$^{rd}$ and 84$^{th}$ bands (highlighted by a red box) at the point $\Gamma$. We find that these two bands display as $k$-relation dispersions along the k$_z$ direction and $k^2$-relation dispersions in the other directions around the point $\Gamma$, which have the Chern numbers of $\pm{2}$ as shown in Fig. 1(c). We also find that twofold degenerate bands along the high-symmetry pathes X-M, Z-A-R, and M-R can form the three nodal surfaces (also called as nodal walls).

In what follows, we tend to discuss the topological features and the corresponding surface states of a CTIWP, which is formed by the 83$^{rd}$ and 84$^{th}$ bands in Fig. 2. We find that three-nodal surfaces ($k_x=\pm\pi$, $k_y=\pm\pi$ and $k_z=\pm\pi$ planes, red planes) and a charge-two Weyl point (a green point) can coexist in the first BZ in Fig. 2(a). It is noted that to our best knowledge, this topological feature is the first time to uncover in a realistic material example. We use the Wilson-loop method to calculate the charity of this CTIWP with the Chern number of -2 (green dotted lines) in Fig. 2(b). As illustrated in Fig. 2(c), the CTIWP with the charge of -2 at the $\widetilde{\textrm{$\Gamma$}}$ point acts as the isolated ``sink" point of the Berry curvature, confirming that the CTIWP is indeed non-trivial. To grasp further the topological features of the CTIWPs, we also calculate the surface states of K$_2$Mg$_2$O$_3$ in the (001) surface BZ. In the (001) surface, the $\Gamma$ (X) is projected to $\widetilde{\textrm{$\Gamma$}}$ ($\widetilde{\textrm{X}}$) as shown in Fig. 1(b). We can observe that a quadratic Dirac cone appears at $\widetilde{\textrm{$\Gamma$}}$ point along the high-symmetry $\widetilde{\textrm{X}}$-$\widetilde{\textrm{$\Gamma$}}$-$\widetilde{\textrm{Y}}$ in the (001) surface BZ in Fig. 2(d). We calculate the isofrequency surface contours at 18.88 THz in Figs. 2(e) and 2(f). We find that there are no double-helicoid surface states and only show a isolated point in  Figs. 2(e) and 2(f), which are different from the previous reports. We can conclude that that the CTIWPs are proved for the first time to be beyond the no-go theorem in phononic systems.

\begin{figure}
\includegraphics[width=3.46in]{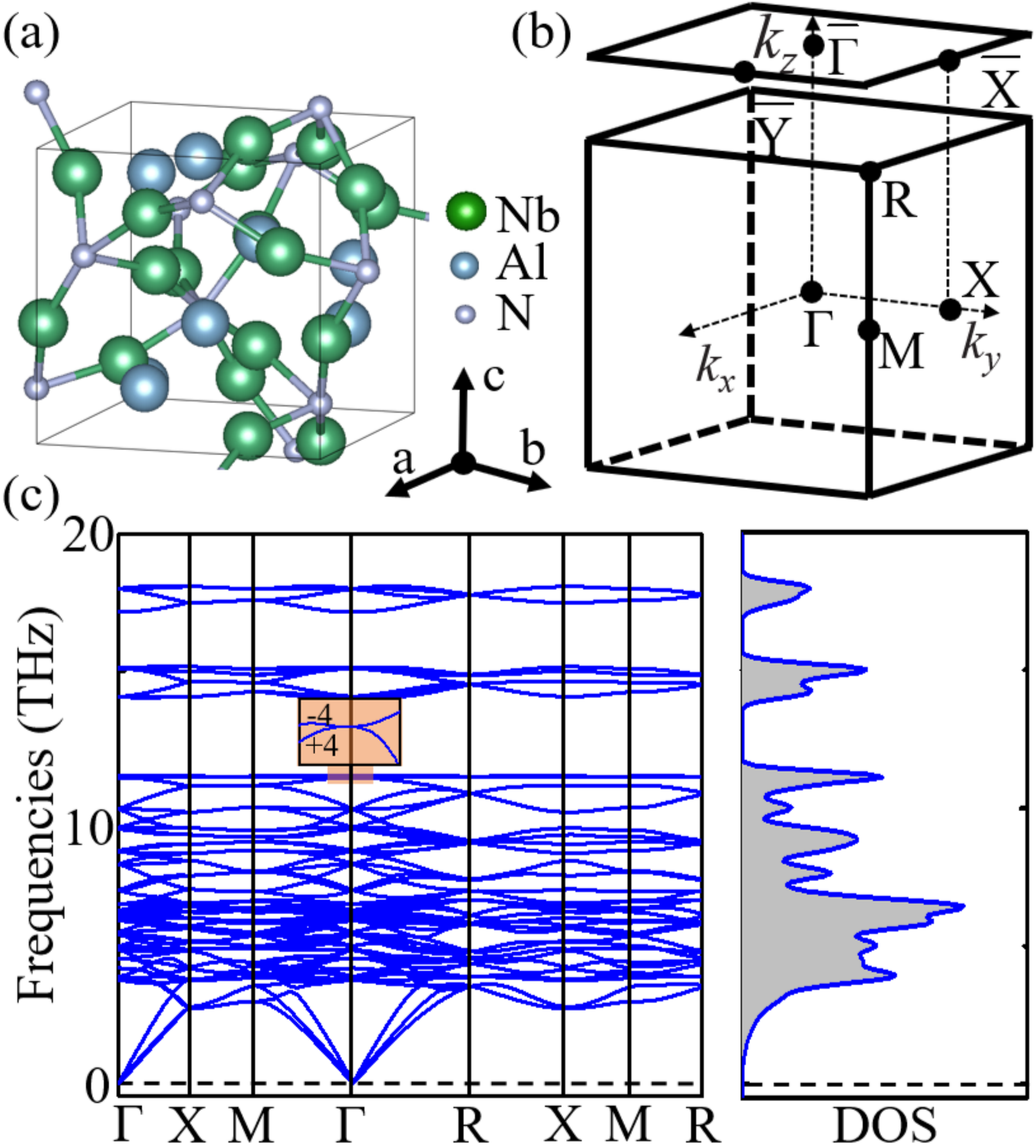}
\caption{The crystalline structures and phonon dispersions of Nb$_3$Al$_2$N in SG 213. (a) Crystal structure of Nb$_3$Al$_2$N in a primitive cell, where the green (light blue and grey) atoms stand for Nb (Al and N). (b) The bulk BZ of Nb$_3$Al$_2$N and the (001) surface BZ. (c) The phononic spectra of Nb$_3$Al$_2$N along the high-symmetry pathes and a red boxes is centered around IWP at $\Gamma$ point, and the Chern numbers ($\pm$4) of some nontrivial phonon branches around the $\Gamma$ point. (d) The phononic density of states (DOS) of Nb$_3$Al$_2$N.}
\end{figure}

\begin{figure*}
\includegraphics[width=17.5 cm]{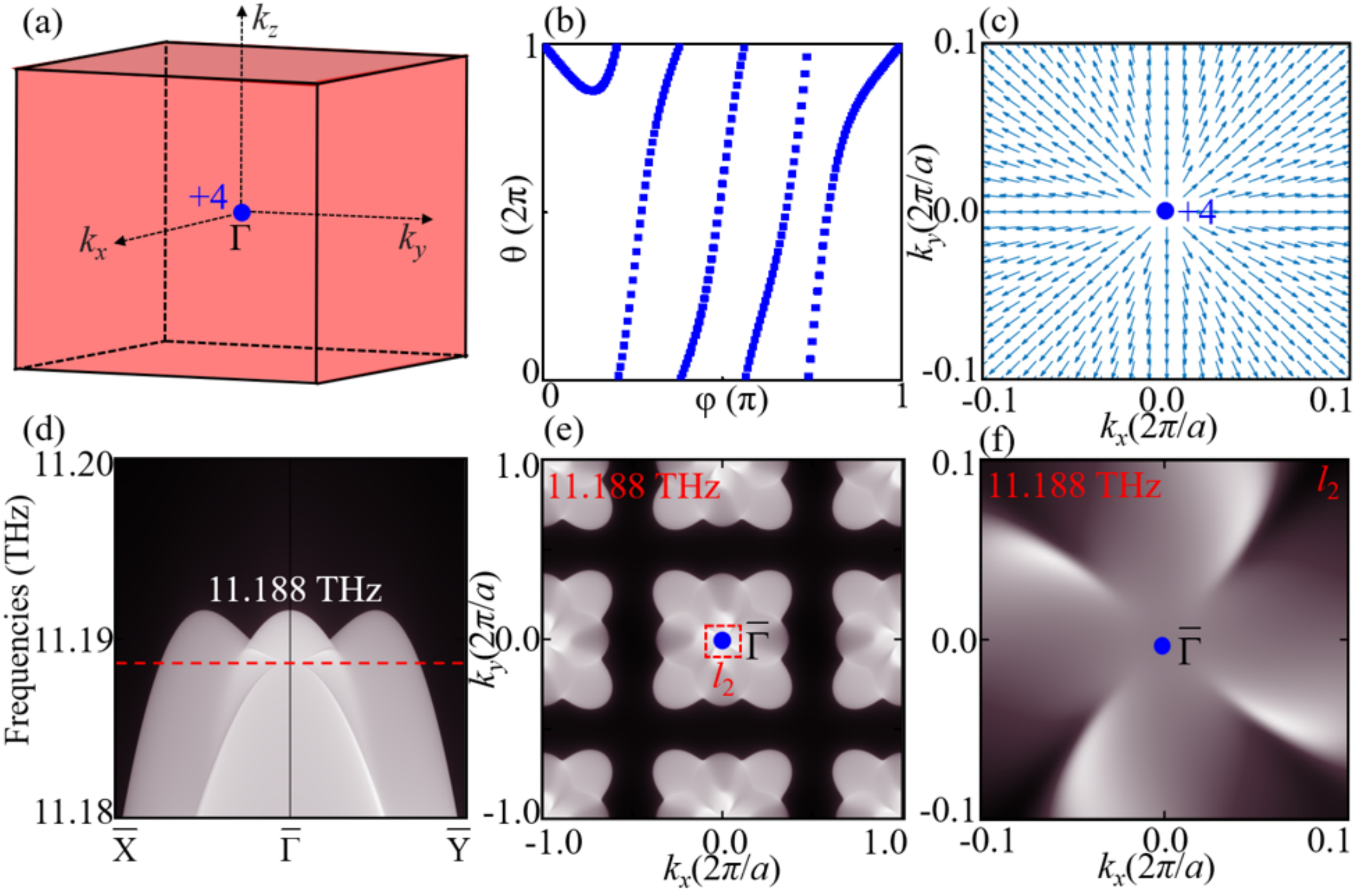}
\caption{The topological features of Nb$_3$Al$_2$N in SG 213. (a) The nodal wall and IWP in the BZ. (b) The evolutions of Wannier centers of the 59th phonon band for $\Gamma$. (c) The distribution of the Berry curvature in the (001) plane. (d) The surface states of Nb$_3$Al$_2$N in the (001) surface BZ along the $\overline{\textrm{X}}$-$\overline{\textrm{$\Gamma$}}$-$\overline{\textrm{Y}}$. (e) and (f) The isofrequency surface contours at 11.188 THz.}
\end{figure*}

{\color{blue}\emph{Topological features and exotic surface states of charge-four IWPs.}} In this part, we will discuss the topological features of charge-four isolated Weyl phonons (CFIWPs). The phononic dispersions and the corresponding PDOS of Nb$_3$Al$_2$N are drawn in Figs. 3(c) and 3(d), which is thermodynamically stable because of no imaginary frequencies in phononic bands. From the phononic dispersions, one can see that there two band crossings are located at the point $\Gamma$ and along the high-symmetry pathes X-M, and R-X-M-R in the optical branches. We mainly discuss a perfect CFIWP which is formed by the 59$^{th}$ and 60$^{th}$ bands (highlighted by a red box) at the point $\Gamma$. We find that these two bands display as $k^3$-relation dispersions along the [111] direction and $k^2$-relation dispersions in the other directions around the point $\Gamma$, which have the Chern numbers of $\pm{4}$ as shown in Fig. 3(c). We also find that the two-fold degenerate bands along the high-symmetry pathes X-M and R-X-M-R can form the three nodal surfaces or nodal walls.

Next, we will discuss the topological features and surface states of a CFIWP, which is formed by the 59$^{th}$ and 60$^{th}$ phononic bands in Fig. 4.  We find that nodal walls ($k_x=\pm\pi$, $k_y=\pm\pi$ and $k_z=\pm\pi$ planes, red planes) and a charge-four Weyl point (a blue point) can coexist in the first BZ as illustrated in Fig. 4(a). It is noted that this inspiring topological feature is the first time to be uncovered in a realistic material example. Moreover, we use the Wilson-loop method to calculate the charity of this CFIWP with the Chern number of +4 (blue dotted lines) in Fig. 4(b). As illustrated in Fig. 4(c), the CFIWP with the charge of +4 at the $\overline{\textrm{$\Gamma$}}$ point acts as the isolated ``source" point of Berry curvature, confirming further that the CFIWP are topologically non-trivial. To further understand the topological nature of the CTIWPs, we also calculate the surface states of Nb$_3$Al$_2$N in the (001) surface BZ. Noted that the $\Gamma$ (X) is projected to $\overline{\textrm{$\Gamma$}}$ ($\overline{\textrm{X}}$) as shown in Fig. 3(b) in the (001) surface BZ. We can observe the quadratic Dirac cone at $\overline{\textrm{$\Gamma$}}$ point along the high-symmetry $\overline{\textrm{X}}$-$\overline{\textrm{$\Gamma$}}$-$\overline{\textrm{Y}}$ in the (001) surface BZ in Fig. 4(d). We calculate the isofrequency surface contours at 11.188 THz in Fig. 4(e) and 4(f). We find that there are no quadruple-helicoid surface states and only show a isolated point as drawn in Figs. 4(e) and 4(f), which are different from the previous reports. In Fig. 4(f), we find that the isofrequency surface contour tends to form a four leaf clover at 11.188 THz. Tt is also the first time to prove that the CFIWPs are beyond the no-go theorem in phononic systems.

In summary, by performing symmetry analysis in 230 SGs, we demonstrated that there exist a new kind of isolated Weyl phonons (IWPs, beyond the no-go theorem' Weyl phonons), \emph{i.e.}, CTIWPs with the Chern numbers of $\pm(2)$ and CFIWPs with the Chern numbers of $\pm{4}$ in phononic crystal systems, and the all twofold IWPs in the high-symmetry $k$-points in 230 SGs are achieved and listed in Table I. By constructing the $k\cdot p$ model of IWPs, we find that the phononic dispersions of CTIWPs display a $k$-type feature along the $k_z$ directions while a $k^2$-type feature along other directions and CFIWPs display a $k^3$-type feature along the [111] directions while a $k^2$-type feature along other directions. Moreover, we chose two realistic material samples, \emph{i.e}., K$_2$Mg$_2$O$_3$ in SG 96 and Nb$_3$Al$_2$N in SG 213, to confirm further their topologically nontrivial nature. For K$_2$Mg$_2$O$_3$, a CTIWP with the charge of -2 appears between the 83$^{rd}$ and 84$^{th}$ bands at the point $\Gamma$. The corresponding surface phononic dispersions display the quadratic Dirac cone and a isolated Weyl point at the isofrequency surface contours. As for Nb$_3$Al$_2$N, the CFIWPs are located at the points $\Gamma$, and a CFIWP appears between the 59$^{th}$ and 60$^{th}$ bands at the point $\Gamma$, the CFIWP-induced the quadratic Dirac cone and a isolated point are also observed in the surface BZ. We find that the CTIWPs and CFIWPs are beyond the no-go theorem and can not form the surface arcs in the surfaces BZ, which is the first time to uncover this physical law in the phononic systems. Our theoretical results not only uncover a new class of Weyl phonons, but also pave an effective way to search the twofold IWPs in spinless systems.

{\color{blue}\emph{Additional notes:}} The IWPs and the corresponding material samples in other SGs listed in Table I are also studied and given in Supplemental Material {\color{blue}\cite{39}}.

{\color{blue}\emph{Acknowledgements.}}\rule[2.50pt]{0.3cm}{0.02em}  This work is supported by the National Science Foundation of China with Grants No. 11774104, No. U20A2077, and No. 12147113, and project funded by the China Postdoctoral Science Foundation with Grant No. 2021M691149.


\begin{thebibliography}{0}%
\makeatletter
\providecommand \@ifxundefined [1]{%
 \@ifx{#1\undefined}
}%
\providecommand \@ifnum [1]{%
 \ifnum #1\expandafter \@firstoftwo
 \else \expandafter \@secondoftwo
 \fi
}%
\providecommand \@ifx [1]{%
 \ifx #1\expandafter \@firstoftwo
 \else \expandafter \@secondoftwo
 \fi
}%
\providecommand \natexlab [1]{#1}%
\providecommand \enquote  [1]{``#1''}%
\providecommand \bibnamefont  [1]{#1}%
\providecommand \bibfnamefont [1]{#1}%
\providecommand \citenamefont [1]{#1}%
\providecommand \href@noop [0]{\@secondoftwo}%
\providecommand \href [0]{\begingroup \@sanitize@url \@href}%
\providecommand \@href[1]{\@@startlink{#1}\@@href}%
\providecommand \@@href[1]{\endgroup#1\@@endlink}%
\providecommand \@sanitize@url [0]{\catcode `\\12\catcode `\$12\catcode
  `\&12\catcode `\#12\catcode `\^12\catcode `\_12\catcode `\%12\relax}%
\providecommand \@@startlink[1]{}%
\providecommand \@@endlink[0]{}%
\providecommand \url  [0]{\begingroup\@sanitize@url \@url }%
\providecommand \@url [1]{\endgroup\@href {#1}{\urlprefix }}%
\providecommand \urlprefix  [0]{URL }%
\providecommand \Eprint [0]{\href }%
\providecommand \doibase [0]{http://dx.doi.org/}%
\providecommand \selectlanguage [0]{\@gobble}%
\providecommand \bibinfo  [0]{\@secondoftwo}%
\providecommand \bibfield  [0]{\@secondoftwo}%
\providecommand \translation [1]{[#1]}%
\providecommand \BibitemOpen [0]{}%
\providecommand \bibitemStop [0]{}%
\providecommand \bibitemNoStop [0]{.\EOS\space}%
\providecommand \EOS [0]{\spacefactor3000\relax}%
\providecommand \BibitemShut  [1]{\csname bibitem#1\endcsname}%
\let\auto@bib@innerbib\@empty
\end{thebibliography}%


\begin{thebibliography}{apssamp}

\bibitem{1} H. Weyl, Gravitation and the electron, {\color{blue}Proc. Natl. Acad. Sci. USA \textbf{15}, 323 (1929)}.

\bibitem{2} B. Bradlyn, J. Cano, Z. Wang, M. G. Vergniory, C. Felser, R. J. Cava, and B. A. Bernevig, Begyond Dirac and Weyl ferimions: unconvertianl quasiparticles in conventional crystals, {\color{blue}Science \textbf{353}, 6299 (2016)}.

\bibitem{3} S.-M. Huang, S.-Y. Xu, I. Belopolski, C.-C. Lee, G. Chang, T.-R. Chang, B. Wang, N. Alidoust, G. Bian, M. Neupane, D. Sanchez, H. Zheng, H.-T. Jeng, A. Bansil, T. Neupert, H. Lin, and M. Z. Hasan, New type of Weyl semimetal with quadratic double Weyl Fermions, {\color{blue}Proc. Nat. Aca. Sci. \textbf{113}, 1180 (2016)}.

\bibitem{4} X. Huang, L. Zhang, Y. Long, P. Wang, D. Chen, Z. Yang, H. Liang, M. Xue, H. Weng, Z. Fang, X. Dai, and G. Chen, Obversation of the chiral-anomaly-induced negative magnetoresistance in 3D Weyl semimetal TaAs, {\color{blue}Phys. Rev. X \textbf{5}, 031023 (2015)}.

\bibitem{5} T. T. Zhang, R. Takahashi, C. Fang, and S. Murakami, Twofold quadruple Weyl nodes in chiral cubic crystals, {\color{blue}Phys. Rev. B \textbf{102}, 125148 (2020)}.

\bibitem{6} Z. M. Yu, Z. Zhang, G. B. Liu, W. Wu, X. P. Li, R. W. Zhang, S. A. Yang, and Y. Yao, Encyclopedia of emergent particles in three-dimensional crystals. Preprint at {\color{blue}arXiv: 2102.01517 (2021)}.

\bibitem{7} L. Zhang and Q. Niu, Chiral phonons at high-symmetry points in monolayer hexagonal lattices, {\color{blue}Phys. Rev. Lett. \textbf{115}, 115502 (2015)}.

\bibitem{8} G. Q. Chang, B. J. Wieder, F. Schindler, D. S. Sanchez, I. Belopolski, S. M. Huang, B. Singh, D. Wu, T. R. Chang, T. Neupert, S.-Y. Xu, H. Lin, and M. Z. Hassan, Topological quantum properties of chiral crystals, {\color{blue}Nat. Mater. \textbf{17}, 978985 (2018)}.

\bibitem{9} C. J. Bradley and A. P. Cracknell, The mathematical theory of symmetry in solids: representation theory for point groups and space groups, {\color{blue}Oxford University Press (2009)}.

\bibitem{10} B. Q. Lv, N. Xu, H. M. Weng, J. Z. Ma, P. Richard, X. C. Huang, L. X. Zhao, G. F. Chen, C. E. Matt, F. Bisti, V. N. Strocov, J. Mesot, Z. Fang, X. Dai, T. Qian, M. Shi, and H. Ding, Observation of Weyl nodes in TaAs, {\color{blue}Nat. Phys. \textbf{11}, 724 (2015)}.

\bibitem{11} S. Y. Xu, I. Belopolski, N. Alidoust, M. Neupane, G. Bian, C. L. Zhang, R. Sankar, G. Q. Chang, Z. J. Yuan, C. C. Lee, S. M. Huang, H. Zheng, J. Ma, D. S. Sanchez, B. K. Wang, A. Bansil, F. C. Chou, P. P. Shibayev, H. Lin, S. Jia, and M. Z. Hasan, Discovery of a Weyl fermion semimental and topological Fermi arcs, {\color{blue}Science \textbf{349}, 613 (2015)}.

\bibitem{12} T.-T. Zhang, Z. D. Song, A. Alexandradinata, H.-M. Weng, C. Fang, L. Lu, and Z. Fang, Double-weyl phonons in transition-metal monosilicides, {\color{blue}Phys. Rev. Lett. \textbf{120}, 016401 (2018)}.

\bibitem{13} H. Miao, T.-T. Zhang, L. Wang, D. Meyers, A.-H. Said, Y.-L. Wang, Y.-G. Shi, H.-M. Weng, Z. Fang, and M.-P.-M. Dean, Observation of double Weyl phonons in parity-breaking FeSi, {\color{blue}Phys. Rev. Lett. \textbf{121}, 035302 (2018)}.

\bibitem{14} Q.-B. Liu, Z. Wang, and H.-H. Fu, Charge-four Weyl phonons, {\color{blue}Phys. Rev. B \textbf{103}, L161303 (2021)}.

\bibitem{15} Q.-B. Liu, Y. Qian, H.-H. Fu, and Z. Wang, Symmetry-enforced Weyl phonons, {\color{blue}npj Comput. Mater. \textbf{6}, 95 (2020)}.

\bibitem{16} L. Lu, Z.- Y. Wang, D.-X. Ye, L.- X. Ran, L. Fu, J. D. Joannopoulos, and M. Soljaci, Experimental observation of Weyl points, {\color{blue}Science \textbf{349}, 622 (2015)}.

\bibitem{17} Q.-B. Liu, H.-H. Fu, and R. Wu, Topological phononic nodal hexahedron net and nodal links in the high-pressure phase of the semiconductor CuCl, {\color{blue}Phys. Rev. B \textbf{104}, 045409 (2021)}

\bibitem{18} Z.-J. Chen, R. Wang, B.-W. Xia, B.-B. Zheng, Y.-J. Jin, Yu-Jun Zhao, and H. Xu, Three-Dimensional Dirac Phonons with Inversion Symmetry, {\color{blue}Phys. Rev. Lett. \textbf{126}, 185301 (2021)}

\bibitem{19} Q.-B. Liu, H.-H. Fu, G. Xu, R. Yu and R. Wu, Categories of phononic topological Weyl open nodal lines and a potential material candidate: Rb$_2$Sn$_2$O$_3$, {\color{blue}J. Phys. Chem. Lett. \textbf{10}, 4045 (2019)}

\bibitem{19_1} Q. Xie, J. Li, S. Ullah, R. Li, L. Wang, D. Li, Y. Li, S. Yunoki, and X.-Q. Chen, Phononic Weyl points and one-way topologically protected nontrivial phononic surface arc states in noncentrosymmetric WC-type materials, {\color{blue}Phys. Rev. B. \textbf{99}, 174306 (2019)}.

\bibitem{19_2} E. Prodan and C. Prodan, Topological phonon modes and their role in dynamic instability of microtubules, {\color{blue}Phys. Rev. Lett. \textbf{103}, 248101 (2009)}.

\bibitem{19_3} C. He, X. Ni, H. Ge, X.-C. Sun, Y.-B. Chen, M.-H. Lu, X.-P. Liu, and Y.-F. Chen, Acoustic topological insulator and robust one-way sound transport, {\color{blue}Nat. Phys. \textbf{12}, 1124 (2016)}.

\bibitem{19_4} G. W Winkler, S. Singh, and A. A. Soluyanov, Topological phononics: from fundamental models to real materials, {\color{blue}Adv. Funct. Mater.  \textbf{30}, 1904784 (2020)}.

\bibitem{19_5} J. Li, Q. Xie, J. Liu, R. Li, Liu, L. Wang, D. Li, Y. Li, and X.-Q. Chen, Phononic Weyl nodal straight lines in MgB$_2$, {\color{blue}Phys. Rev. B. \textbf{101}, 024301 (2020)}.

\bibitem{20} Y. Z. Liu, Y. Xu, S.-C. Zhang, and W. H. Duan, Model for topological phononics and phonon diode, {\color{blue}Phys. Rev. B \textbf{96}, 064106 (2017)}.

\bibitem{21} L. F. Zhang, J. Ren, J. S. Wang, and B. W. Li, Topological nature of the phonon Hall effect, {\color{blue}Phys. Rev. Lett. \textbf{105}, 225901 (2010)}.

\bibitem{22} R.- H. Li, H. Ma, X.- Y. Cheng, S. L. Wang, D.-Z. Li, Z.- Y. Zhang, Y.-Y. Li, and X.-Q. Chen, Dirac Node Lines in Pure Alkali Earth Metals, {\color{blue}Phys. Rev. Lett. \textbf{117}, 096401 (2016)}.

\bibitem{23} W.-Y. Deng, J.-Y. Lu, F. Li, X.-Q. Huang, M. Yan, J.-H. Ma, and Z.-Y. Liu, Nodal rings and drumhead surface states in phononic crystals, {\color{blue}Nat. Commun. \textbf{10}, 1769 (2019)}.

\bibitem{24} Y.- J. Jin, Z.- J. Chen, B.- W. Xia, Y.- J. Zhao, R. Wang, and H. Xu, Ideal intersecting nodal-ring phonons in bcc C8, {\color{blue}Phys. Rev. B. \textbf{98}, 220103(R) (2018)}.

\bibitem{25} B. Zheng, B. Xia, R. Wang, Z. Chen, J. Zhao, Y. Zhao, and H. Xu, Ideal type-III nodal-ring phonons, {\color{blue}Phys. Rev. B \textbf{101}, 100303(R) (2020)}.

\bibitem{26} G. Q. Chang, S.-Y. Xu, X. T. Zhou, S.-M. Huang, B. Singh, B. K. Wang, I. Belopolski, J. X. Yin, S. T. Zhang, A. Bansil, H. Lin, and M. Z. Hasan, Topological Hopf and Chain Link Semimetal States and Their Application to Co2MnGa, {\color{blue}Phys. Rev. Lett. \textbf{119}, 156401 (2017)}.

\bibitem{26_1}  D.-F. Shao, E. Y. Tsymbal, S.-H. Zhang and X. Dang, Tunable two-dimensional Dirac nodal nets, {\color{blue}Phys. Rev. B \textbf{98}, 161104(R) (2018)}.

\bibitem{27} R. Bi, Z. Yan, L. Lu, and Z. Wang, Nodal-knot Semimetals, {\color{blue}Phys. Rev. B \textbf{96}, 201305(R) (2017)}.

\bibitem{27_1} C.H.Lee, A. Sutrisno, T. Hofmann,  et al. Imaging nodal knots in momentum space through topolectrical circuits. {\color{blue}Nat. Commun. \textbf{11}, 4385 (2020)}.

\bibitem{27_2} R. Bi, Z. Yan, L. Lu, and Z. Wang, Nodal-knot Semimetals, {\color{blue}Phys. Rev. B  \textbf{96}, 201305(R) (2017)}.

\bibitem{28} Y. Sun, Y. Zhang, C.-X. Liu, C. Felser, and B. H. Yan, Dirac nodal lines and induced spin Hall effect in metallic rutile oxides, {\color{blue}Phys. Rev. B. \textbf{95}, 235104 (2017)}.

\bibitem{29} B. W. Xia, R. Wang, Z. J. Chen, Y. J. Zhao, and H. Xu, Symmetry-protected ideal type-II Weyl phonons in CdTe, {\color{blue}Phys. Rev. Lett. \textbf{123}, 065501 (2019)}.


\bibitem{30} H. Geoffroy, F. Chris, E. Virginie, J. Anubhav, and C. Gerbrand, Data mined ionic substitutions for the discovery of new compounds, {\color{blue}Inorg. Chem. \textbf{50}, 656 (2011)}.



\bibitem{31} W. Kohn and L. J. Sham, Self-consistent equations including exchange and correlation effects, {\color{blue}Phys. Rev. B \textbf{136}, A1133 (1964)}.


\bibitem{32} P. E. Bl\"{o}chl, Projector augmentad-wave method, {\color{blue}Phys. Rev. B \textbf{50}, 17953 (1994)}.

\bibitem{33} J. P. Perdew, K. Burke, and M. Ernzerhof, Generalized gradient approximation made simple, {\color{blue}Phys. Rev. Lett. \textbf{77}, 3865 (1996)}.

\bibitem{34} A. Togo and I. Tanaka, First principles phonon calculations in materials science, {\color{blue}Scr. Mater. \textbf{108}, 1 (2015)}.

\bibitem{35} Q.-S. Wu, S.-N. Zhang, H.-F. Song, M. Troyer, and A. A. Soluyanov, WannierTools: an open-source software package for novel topological materials, {\color{blue}Comput. Phys. Commun. \textbf{243}, 110 (2019)}.

\bibitem{36} M. P. Lopez Sancho, J. M. Lopez Sancho, J. M. L. Sancho, and J. Rubio, Highly convergent schemes for the calculations of bulk and surface Green functions, {\color{blue}J. Phys. F: Metal Phys., \textbf{15}, 851 (1985)}.

\bibitem{37} R. Yu, X. L. Qi, A. Bernevig, Z.Fang \& Dai, X. Equivalent expression of $Z_2$ topological invariant for band insulators using the non-abelian Berry connection. {\color{blue}Phys. Rev. B \textbf{84}, 075119 (2011)}.

\bibitem{38} J. C. Gao, Q. S. Wu,  C. Persson, and Z. Wang, Irvsp: To obtain irreducible representations of electronic states in the VASP, {\color{blue}Comput. Phys. Commun., \textbf{261}, 107760 (2021)}.


\bibitem{39} {\color{blue}See Supplemental Material at http://link.aps.org/ supplemetnal/xxxxxx}].







\end{thebibliography}
\end{document}